\documentclass[12pt]{article}
\usepackage[dvips]{graphicx}

\def\be{\begin{equation}}
\def\ee{\end{equation}}
\def\bea{\begin{eqnarray}}
\def\eea{\end{eqnarray}}
\def\fixit#1{}

\def\href#1#2{#2}

\def\lbldef#1#2{\expandafter\gdef\csname #1\endcsname {#2}}

\begin{document}
\baselineskip=16pt \pagestyle{plain} \setcounter{page}{1}
%\renewcommand{\thefootnote}{\fnsymbol{footnote}}
%--------+---------+---------+---------+---------+---------+---------+
%Title page

\begin{titlepage}

\begin{flushright}
hep-th/0611135 \\
PUPT-2216
\end{flushright}
\vfil

\begin{center}
{\huge A Test of the AdS/CFT Correspondence\\ Using High-Spin Operators
}
\end{center}

\vfil
\begin{center}
{\large M. K. Benna$^a$ \footnote{E-mail: {\tt
mbenna@princeton.edu}}, S. Benvenuti$^a$ \footnote{E-mail: \tt
sbenvenu@princeton.edu}, I. R. Klebanov$^{a,b}$ \footnote{E-mail:
\tt klebanov@princeton.edu}, A. Scardicchio$^{a,b}$
\footnote{E-mail: \tt ascardic@princeton.edu} }\end{center}

\vfil

$^a$Joseph Henry Laboratories, Princeton University,
 Princeton, NJ  08544

$^b$Princeton Center for Theoretical Physics, Princeton
University, Princeton, NJ  08544 \vfil

\begin{center}
{\large Abstract}
\end{center}

\noindent In two remarkable recent papers, hep-th/0610248 and
hep-th/0610251, the complete planar perturbative expansion was
proposed for the universal function of the coupling, $f(g)$,
appearing in the dimensions of high-spin operators of the ${\cal
N}=4$ SYM theory. We study numerically the integral equation
derived in hep-th/0610251, which implements a resummation of the
perturbative expansion, and find a smooth function that approaches
the asymptotic form predicted by string theory. In fact, the two
leading terms at strong coupling match with high accuracy the
results obtained for the semiclassical folded string spinning in
$AdS_5$. This constitutes a remarkable confirmation of the AdS/CFT
correspondence for high-spin operators, and equivalently for the
cusp anomaly of a Wilson loop. We also make a numerical prediction for
the third term in the strong coupling series. \vfil
\begin{flushleft}
I.R. Klebanov dedicates this paper to the memory of his
brother-in-law, Gordon E. Kato.
\end{flushleft}
\end{titlepage}
\newpage
\renewcommand{\thefootnote}{\arabic{footnote}}
\setcounter{footnote}{0}
\renewcommand{\baselinestretch}{1.2}  %looks better
%\tableofcontents
%--------+---------+---------+---------+---------+---------+---------+
%Body

\section{Introduction}
%\label{Introduction}

The dimensions of high-spin operators are important observables in
gauge theories. It is well-known that the anomalous dimension of a
twist-2 operator grows logarithmically for large spin $S$,
\begin{equation} \label{Universal}
\Delta-S = f(g) \ln S\ + O(S^0), \qquad g ={\sqrt{g_{YM}^2 N}\over
4\pi}\ .
\end{equation}
This was demonstrated early on at 1-loop order \cite{GrossWilczek}
and at two loops \cite{Floratos} where a cancellation of $\ln^3 S$
terms occurs.  There are solid arguments that (\ref{Universal})
holds to all orders in perturbation theory
\cite{Korchemsky,Sterman}, and that it also applies
to high-spin operators of twist greater
than two \cite{Belitsky:2006en}.
The function of coupling $f(g)$ also
measures the anomalous dimension of a cusp in a light-like Wilson
loop, and is of definite physical interest in QCD.

There has been significant interest in determining $f(g)$ in the
${\cal N}=4$ SYM theory. This is partly due to the fact that the
AdS/CFT correspondence \cite{AdSCFT} relates the large $g$
behavior of $f$ to the energy of a folded string spinning around
the center of a weakly curved $AdS_5$ space \cite{GKP}. This gives
the prediction that $f(g)\rightarrow 4 g$ at strong coupling. The
same result was obtained from studying the cusp anomaly using
string theory methods \cite{Kruczenski}. Furthermore, the
semi-classical expansion for the spinning string energy predicts
the following correction \cite{FT}:
\begin{equation} \label{AdSpredict}
f(g) = 4 g - {3\ln 2\over \pi} + O(1/g).
\end{equation}
It is of obvious interest to confirm these explicit predictions of
string theory using extrapolation of the perturbative expansion
for $f(g)$ provided by the gauge theory.

Explicit perturbative calculations are quite formidable, and until
recently were available only up to 3-loop order
\cite{Kotikov,Bern}:
\begin{equation}
f(g) = 8 g^2 -{8\over 3} \pi^2 g^4 + {88\over 45} \pi^4 g^6 +
O(g^8).
\end{equation}
Kotikov, Lipatov, Onishchenko and Velizhanin \cite{Kotikov}
extracted the ${\cal N}=4$ answer from the QCD calculation of
\cite{Moch} using their proposed transcendentality principle
stating that each expansion coefficient has terms of the same
degree of transcendentality.

Recently, the methods of integrability in AdS/CFT\footnote
{For earlier work on integrability in gauge theories, see
\cite{Lipatov,Faddeev,Braun}.} \cite{Minahan,Beisert:2004ry},
%(for reviews
%see \cite{Beisert:2004ry, Belitsky:2004cz}),
prompted in part by
\cite{BMN,GKP}, have led to dramatic progress in
studying the weak coupling expansion. In the beautiful paper by
Beisert, Eden and Staudacher \cite{BES}, which followed closely
the important earlier work in \cite{Eden,BHL}, an integral
equation that determines
$f(g)$ was proposed, yielding an expansion of $f(g)$
to an arbitrary desired order. The expansion coefficients obey the
KLOV transcendentality principle. In an independent remarkable
paper by Bern, Czakon, Dixon, Kosower and Smirnov \cite{BCDKS}, an
explicit calculation led to a value of the 4-loop term,
\begin{equation}
-16 \left ({73\over 630} \pi^6 + 4\zeta (3)^2 \right ) g^8 \
,\end{equation}
which agrees with the idea advanced in
\cite{BCDKS,BES} that the exact expansion of $f(g)$ is related to
that found in \cite{Eden} simply by multiplying each
$\zeta$-function of an odd argument by an $i$, $\zeta(2n+1)\rightarrow
i \zeta(2n+1)$. The integral
equation of $\cite{BES}$ generates precisely this perturbative
expansion for $f(g)$.

A crucial property of the integral equation proposed in
\cite{BES} is that it is related through integrability to the
``dressing phase'' in the magnon S-matrix, whose general form was
deduced in \cite{AFS,BK}. In \cite{BES} a perturbative expansion
of the phase was given, which starts at the 4-loop order, and at
strong coupling coincides with the earlier results from string
theory \cite{AFS,HL,Freyhult:2006vr,HM,BHL}.
An important requirement of crossing
symmetry \cite{Janik} is satisfied by this phase, and it also
satisfies the KLOV transcendentality priciple. Therefore, this
phase is very likely to describe the exact magnon S-matrix at any
coupling \cite{BES}, which constitutes remarkable progress in the
understanding of the ${\cal N}=4$ SYM theory, and of the AdS/CFT
correspondence.

The papers \cite{BES,BCDKS} thoroughly studied the perturbative
expansion of $f(g)$ which follows from the integral equation.
Although the expansion has a finite radius of convergence, as is
customary in certain planar theories (see, for example,
\cite{GT}), it is expected to determine the function completely.
Solving the integral equation of \cite{BES} is an efficient tool
for attacking this problem. In this paper we solve the integral
equation numerically at intermediate coupling, and show that
$f(g)$ is a smooth function that approaches the asymptotic form
(\ref{AdSpredict}) predicted by string theory for $g> 1$. The two
leading strong coupling terms match those in (\ref{AdSpredict})
with high accuracy. This constitutes a remarkable confirmation of
the AdS/CFT correspondence for this non-supersymmetric observable.

The structure of the paper is as follows. The integral equation
of \cite{BES} is reviewed and solved numerically in section 2.
An interpretation
of these results and their implications for the AdS/CFT
correspondence are given in section 3.

\section{Numerical study of the integral equation}

The cusp anomalous dimension $f(g)$ can be written as
\cite{Eden,BES,Belitsky:2006wg}
\be
f(g)=16g^2\hat\sigma(0),
\ee
where $\hat \sigma(t)$ obeys the integral equation
\be
\hat\sigma(t)=\frac{t}{e^t-1}\left(K(2gt,0)-4g^2\int_0^\infty dt'
K(2gt,2gt')\hat\sigma(t')\right),
\ee
with the kernel given by \cite{BES}
\be \label{corrkern} K(t,t') = K^{(m)}(t,t') + 2
K^{(c)}(t,t').
\ee
The main scattering kernel $K^{(m)}$ of \cite{Eden} is
\be
K^{(m)}(t,t')=\frac{J_1(t)J_0(t')-J_0(t)J_1(t')}{t-t'},
\ee
and the dressing kernel $K^{(c)}$ is defined as the convolution
\be
K^{(c)}(t,t') = 4g^2 \int_0^\infty dt'' K_1(t,2gt'')
\frac{t''}{e^{t''}-1} K_0(2gt'',t'),
\ee
where $K_0$ and $K_1$ denote the parts of the kernel that are even
and odd, respectively, under change of sign of $t$ and $t'$:
\bea
K_0(t,t') &=& \frac{t J_1(t) J_0(t') - t' J_0(t) J_1(t') }{t^2-t'^2} = {2\over t\, t'}\sum_{n=1}^{\infty} (2n-1) J_{2n-1}(t)J_{2n-1}(t'), \\
K_1(t,t') &=& \frac{t' J_1(t) J_0(t') - t J_0(t) J_1(t')
}{t^2-t'^2} = {2\over t\, t'}\sum_{n=1}^{\infty} (2n)
J_{2n}(t)J_{2n}(t').
\eea
We find it useful to introduce the function \be s(t) = {e^t-1
\over t}\,\hat{\sigma}(t), \ee in terms of which the integral
equation becomes
\be
\label{inteqs}
s(t) = K(2gt,0) - 4g^2\int_0^\infty dt'
K(2gt,2gt')\frac{t'}{e^{t'}-1}s(t').
\ee
Again, $f(g)=16g^2s(0)$.

Both $K^{(m)}$ and $K^{c)}$ can conveniently be expanded as sums
of products of functions of $t$ and functions of $t'$:
\be \label{Km}
K^{(m)}(t,t') = K_0(t,t') + K_1(t,t') = {2\over t\,t'}
\sum_{n=1}^{\infty} n J_n(t)J_n(t'),
\ee
and
\be \label{Kc}
K^{(c)}(t,t') = \sum_{n=1}^{\infty}\sum_{m=1}^{\infty} {8 n
(2m-1)\over t\,t'}Z_{2n,2m-1} J_{2n}(t)J_{2m-1}(t').
\ee

This suggests writing the solution in terms of linearly
independent functions
%$J_n(t)/t$
as
\be
\label{s-expa}
s(t)=\sum_{n\geq 1}s_n\frac{J_n(2gt)}{2gt},
\ee
so that the integral equation becomes a matrix equation for the
coefficients $s_n$. The desired function $f(g)$ is now
\be
f(g)=8g^2s_1.
\ee

It is convenient to define the matrix $Z_{mn}$ as
\be
Z_{mn} \equiv \int_0^{\infty}dt {J_m(2gt) J_n(2gt) \over t
(e^t -1)}.
\ee
Using the representations (\ref{Km}) and (\ref{Kc}) of the kernels
and (\ref{s-expa}) for $s(t)$, the integral equation above is now
of the schematic form
\be \label{expansion}
s_n=h_n-\sum_{m\geq 1}\left(K^{(m)}_{nm}+2K^{(c)}_{nm}\right)s_m,
\ee
whose solution is
\be
s=\frac{1}{1+K^{(m)}+2K^{(c)}}h.
\ee
The matrices are
\bea
K^{(m)}_{nm}&=&2(NZ)_{nm},\\
K^{(c)}_{nm}&=&2(CZ)_{nm},\\
C_{nm}&=&2(PNZQN)_{nm},
\eea
where $Q={\rm diag}(1,0,1,0,...)$, $P={\rm diag}(0,1,0,1,...)$,
$N={\rm diag}(1,2,3,...)$ and the vector $h$ can be written as
$h=(1+2C)e^\mathrm{T}$, where $e=(1,0,0,...)$. The crucial point for the
numerics to work is that the matrix elements of $Z$ decay
sufficiently fast with increasing $m,n$ (they decay like $e^{-{\rm
max}(m,n)/g}$). For intermediate $g$ (say $g<20$) we can work with moderate size $d$ by $d$ matrices, where $d$ does not have to be much larger than $g$. The integrals in $Z_{nm}$ can be
obtained numerically without much effort and so we can solve for
the $s_n$. We find that the results are stable
with respect to increasing $d$.

Even though at strong coupling all elements of $Z_{nm}$ are of the same
order in $1/g$, those far from the upper left corner are
numerically small (the leading terms in $1/g$ are suppressed by a
factor $[(m-n-1)(m-n+1)(m+n-1)(m +n+1)]^{-1}$ for $m \neq n \pm
1$). This last fact makes the numerics surprisingly convergent
even at large $g$ and moreover gives some hope that the analytic
form of the strong coupling expansion of $f(g)$ could be obtained
from a perturbation theory for the matrix equation.

Therefore, when formulated in terms of the $Z_{mn}$, the problem
becomes amenable to numerical study at all values of the coupling.
We find that the numerical procedure converges rather rapidly, and
truncate the series expansions of $s(t)$ and of the kernel after the
first 30 orders of Bessel functions.

The function $f(g)$ is the lowest curve plotted in Figure 1.
For comparison, we also plot $f_m(g)$ which solves the integral
equation with kernel $K^{(m)}$ \cite{Eden}, and $f_0 (g)$
which solves the integral
equation with kernel $K^{(m)} + K^{(c)}$. Clearly, these functions differ
at strong coupling.
The function $f(g)$ is monotonic and reaches the asymptotic, linear
form quite early, for $g\simeq 1$. We can then study the
asymptotic, large $g$ form easily and compare it with the
prediction from string theory. The best fit result (using the
range $2<g<20$) is
\be
f(g) = (4.000000\pm 0.000001)g - (0.661907 \pm 0.000002) -
\frac{0.0232 \pm 0.0001}{g} + \ldots
\ee
The first two terms are in remarkable agreement with the string
theory result (\ref{AdSpredict}), while the third
term is a numerical prediction for the $1/g$ term in the strong
coupling expansion, which may perhaps be checked one day against a
two-loop string theory calculation.

\begin{figure}
\begin{center}
\includegraphics[width = 0.8\textwidth]{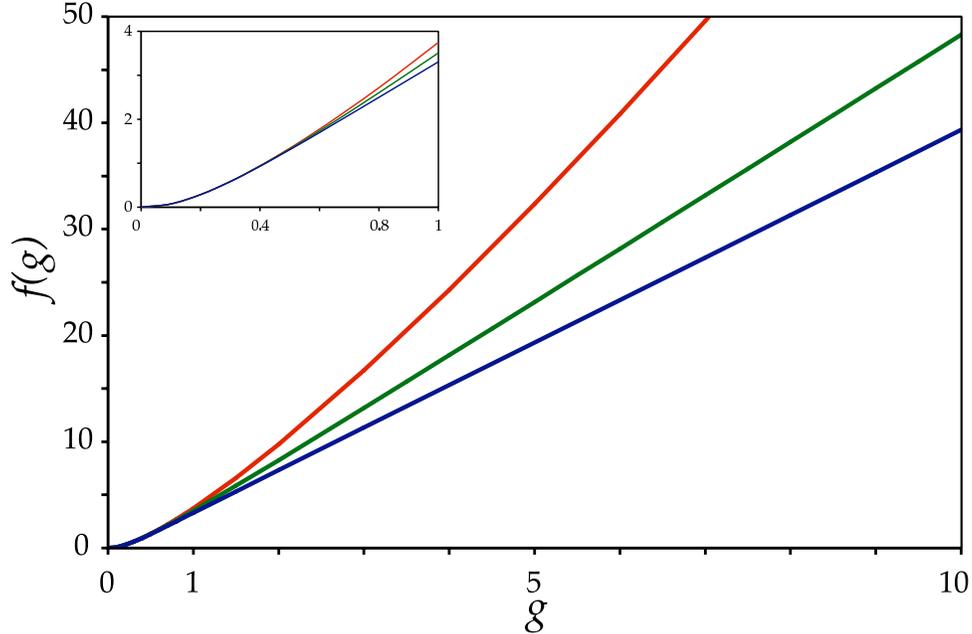}
\end{center}
\caption{Plot of the solutions of the integral equations: $f_m(g)$ for the ES
kernel $K^{(m)}$ (upper curve, red),
$f_0(g)$ for the kernel $K^{(m)} +
K^{(c)}$ (middle curve, green), and $f(g)$ for
 the BES kernel $K^{(m)} +
2K^{(c)}$ (lower curve, blue). Notice the different asymptotic
behaviors. The inset shows the three functions in the crossover
region $0<g<1$.}
\end{figure}

We do not need to restrict the numerical analysis to real values
of $g$; complex values of $g$ are of interest as well.
In \cite{BES} it was argued that the dressing phase has
singularities at $g\approx \pm in/4$, for $n=1,2,3,\ldots$.
Also, their analysis of the small $g$ series shows that there are
square root branch points in $f(g)$ at $g=\pm i/4$.
Perhaps, this is related to the
cuts in the giant-magnon dispersion relations
\cite{Beisert:2005tm, BDS, Berenstein:2005jq, Santambrogio:2002sb, HM},
for momenta close to $\pi$.
Our numerical results indeed indicate
branch points at $g\approx \pm i/4, \pm i/2$
with exponent $1/2$. Beyond that we observe oscillations of both the real
and imaginary parts of $f(g)$ for nearly imaginary $g$.
Further work is needed to elucidate the analytical structure of
$f(g)$.

\section{Discussion}

A very satisfying result of this paper is that the
BES integral equation yields a smooth universal function
$f(g)$ whose strong coupling expansion is in excellent numerical
agreement with the spinning string predictions of \cite{GKP,FT}.
This provides a highly non-trivial confirmation of the AdS/CFT
correspondence.

The agreement of this strong coupling expansion was anticipated in
\cite{BES} based on a similar agreement of the dressing phase.
However, some concerns about this argument were raised in
\cite{BCDKS} based on the slow convergence of the numerical
extrapolations. Luckily, our numerical methods employed in solving
the integral equation converge rapidly and produce a smooth
function that approaches the asymptotics (\ref{AdSpredict}). The
cross-over region of $f(g)$ where it changes from
the perturbative to the linear behavior lies right around the
radius of convergence, $g_c=1/4$,
corresponding to $g_{YM}^2 N=\pi^2$. For $N=3$, this would
correspond to $\alpha_s \sim 0.25$.

The qualitative structure of the interpolating function $f(g)$ is
quite similar to that involved in the circular Wilson loop, where
the conjectured exact result \cite{esz,GrossDrukker} is
\begin{equation}\label{WL}
\ln \left ({I_1(4\pi g)\over 2\pi g}\right )\ .
\end{equation}
The function (\ref{WL}) is analytic on the complex plane, with a
series of branch cuts along the imaginary axis, and an essential
singularity at infinity. The function $f(g)$ is also expected
to have an infinite number of branch cuts along the imaginary
axis, and an essential singularity at infinity \cite{BES}.

Let us compare this with the exact anomalous
dimension of a single giant magnon of momentum $p$
\cite{Beisert:2005tm, BDS, Berenstein:2005jq, Santambrogio:2002sb, HM}:
\begin{equation}\label{SM}
-1 + \sqrt{1+16 g^2 \sin^2\left(\frac{p}{2}\right)}\ .
\end{equation}
This function has a single branch cut along the imaginary axis,
going from $g=\frac{i}{4\sin\left(\frac{p}{2}\right)}$ to
infinity, and no essential singularity at infinity.
Observables of the gauge theory are composites of giant magnons
with various momenta $p\in (0,2\pi)$. We thus expect the
anomalous dimension of a generic unprotected operator to have a
superposition of many cuts along the imaginary axis. This
should endow a generic multi-magnon state with an analytic structure
similar to that of $f(g)$.

We found numerically the presence, in $f(g)$, of the first two branch
cuts on the imaginary axis, starting at $g=\pm \frac{ni}{4}$, $n=1,2$.
The first of them, which also occurs for
the giant magnon with maximal momentum
$p=\pi$, agrees with the summation of the perturbative series
\cite{BES}. The full structure of $f(g)$ is expected to contain an
infinite number of branch cuts accumulating at infinity, where an
essential singularity is present.

It is remarkable that the integral equation of \cite{BES} allows
$f(g)$, which is not a BPS quantity, to be solved for. Hopefully, this paves
the way to finding other observables as functions of the
coupling in the planar ${\cal N}=4$ SYM theory.

%%%%%%%%%%%%%%%%%%%%%%%%%%%%%%%%%%%%%%%%%%%%%%%
\section*{Acknowledgements}
%%%%%%%%%%%%%%%%%%%%%%%%%%%%%%%%%%%%%%%%%%%%%%%%%%%%
We thank J. Maldacena, M. Staudacher and A. Tseytlin for useful discussions.
This research was supported in part by the National Science Foundation
under Grant No. PHY-0243680. Any opinions, findings, and
conclusions or recommendations expressed in this material are
those of the authors and do not necessarily reflect the views of
the National Science Foundation.

\begingroup\raggedright\endgroup

\end{document}